\documentclass[twocolumn,showpacs,preprintnumbers,amsmath,amssymb]{revtex4}

\usepackage{graphicx}% Include figure files
\usepackage{dcolumn}% Align table columns on decimal point
\usepackage{bm}% bold math

\begin{document}
{\bf Comment on ``Interaction Effects in Conductivity of Si Inversion Layers at Intermediate Temperatures''}\vspace{3mm}

In a recent Letter \cite{pudalov03}, Pudalov {\it et al}.\ have claimed that they have found an excellent agreement between their experimental data and theory \cite{zala01}. According to them, the anomalous (by an order of magnitude \cite{kravchenko03}) increase of resistance with temperature can be quantitatively described by the theory of small corrections arising from quantum coherent interaction effects. Notably, this claim negates previous publication \cite{brunthaler01} entitled ``{\em Exclusion of Quantum Coherence as the Origin of the 2D Metallic State in High-Mobility Silicon Inversion Layers}'', in which the same samples were studied in the same temperature and density ranges. The purpose of this Comment is to show that when analyzed correctly, the data of Pudalov {\it et al}.\ are not appropriate for accurate comparison with theory and do not allow one to distinguish between the interaction-based theory \cite{zala01} and the traditional screening theory \cite{gold86}.

Despite theory \cite{zala01} yields a linear-in-$T$ correction to {\em conductivity} ($\sigma$), Pudalov {\it et al}.\ choose to analyze the temperature-dependent {\em resistivity} ($\rho$). This serves to mask the fact that there is essentially no linear-in-$T$ interval on their $\sigma(T)$ dependences \cite{temp}. To demonstrate this, in Fig.~1 we plot $\sigma(T)$ recalculated from the published $\rho(T)$ data for two lowest electron densities \cite{lowest} shown in Fig.~1~(b) of Ref.~\cite{pudalov03}. The solid lines correspond to the slopes $d\sigma/dT$, which the authors claim to be in ``excellent agreement'' with their experimental data (circles). The obvious reason for the dramatic discrepancy between presentations of the data in the conductivity and resistivity forms are large changes of $\rho$ with temperature, reaching a factor of two.

Another point of importance is that the data of Ref.~\cite{pudalov03} are not related to the ``{\em anomalous} increase of $\rho$ with temperature'' as they are taken at (relatively) high electron densities $n_s>2\times10^{11}$~cm$^{-2}$, where interaction effects are weak. Metallic corrections to conductivity in this regime have been observed \cite{dolgopolov88} and described by the traditional screening theory \cite{gold86} long time ago. The same theory provides reasonably good fits to the current data from Ref.~\cite{pudalov03} (see, {\it e.g.}, the dashed lines in Fig.~1). Note that at electron densities $\gtrsim5\times10^{11}$~cm$^{-2}$, theory \cite{zala01} can hardly be applied at all to Si inversion layers because surface roughness scattering becomes dominant.

\begin{figure}[ht]\vspace{6mm}
\begin{center}
\scalebox{.56}{\includegraphics{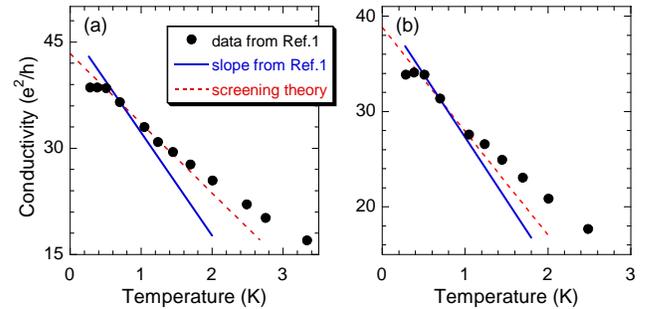}}
\end{center}\vspace{-3mm}
\caption{Conductivity {\it vs}.\ temperature for $n_s=2.46$ (a) and $2.23\times10^{11}$~cm$^{-2}$ (b) along with the slopes $d\sigma/dT$ recalculated from $d\rho/dT$ of Ref.~\protect\cite{pudalov03}. Also shown are the fits to the data using the traditional screening theory \protect\cite{gold86} with the effective mass values taken from Ref.~\cite{shashkin}.}\vspace{-5mm}
\end{figure}

Summarizing, the comparison performed by Pudalov {\it et al}.\ \cite{pudalov03} between experimental data (including those for the magnetoresistance) and theory \cite{zala01} is not valid. Moreover, their data do not support the interaction-related origin of the metallic $\rho(T)$ at the relatively high electron densities used once both
theories \cite{zala01,gold86} are treated on equal footing. As follows from accurate measurements in best samples \cite{shashkin,kravchenko03}, the interactions become important at lower densities close to the metal-insulator transition, where they lead to a strong renormalization of the effective mass which corresponds to strongly temperature-dependent metallic resistivity.

AAS and VTD gratefully acknowledge financial support by the RFBR and the programs ``Nanostructures'' and ``Mesoscopics'' from the Russian Ministry of Sciences. SVK is supported by NSF grant DMR-0129652 and the Sloan Foundation.\vspace{5mm}\\
A.~A. Shashkin and V.~T. Dolgopolov\\
{\it Institute of Solid State Physics, Chernogolovka, Moscow District
142432, Russia}\vspace{2mm}\\
S.~V. Kravchenko\\
{\it Physics Department, Northeastern University, Boston,
Massachusetts 02115, U.S.A.}


\begin{thebibliography}{11}
\bibitem{lowest} The higher-density data are presented in Ref.~\cite{pudalov03} in such a format that it is impossible to judge the quality of fits.
\end{thebibliography}
\end{document}